\documentclass[a4j]{article}
\usepackage[dvips]{graphicx}
\begin{document}

\title{ Multiplicity of Limit Cycle Attractors in 
Coupled Heteroclinic Cycles
}

\author{
Masashi {\sc Tachikawa}\footnote{
E-mail address: mtach@allegro.phys.nagoya-u.ac.jp}\\
{\it Department of Physics, Nagoya University, Nagoya 464-8602, JAPAN}
}
\date{}

\maketitle

\begin{abstract}
A square lattice distribution
of coupled oscillators that have heteroclinic cycle attractors
is studied. In this system, we find 
a novel type of patterns that is spatially
disordered and periodic in time. 
These patterns are limit cycle attractors in the ambient phase space 
(i.e. not chaotic) and many limit cycles exist dividing 
the phase space as their basins. 
The patterns are constructed with a local law of difference 
of phases between the oscillators. The number of patterns 
grows exponentially with increasing of the number of oscillators. 
\end{abstract}

In the recent decades, coupled oscillators have attracted
much attention. 
They are adopted as models for rhythmic or chaotic behaver 
of biological and other complex systems\cite{rf:W80,rf:K84,rf:WCS96}. 
Moreover, coupled oscillators themselves are thought as 
the important classes to investigate 
in high dimensional dynamical systems\cite{rf:K84,rf:D92,rf:MS90}.
For the analytical simplicity, 
limit cycles of a normal form type or phase oscillators 
have been frequently studied to replace detailed dynamics. 
While they have led to much development in understanding of 
high dimensional dynamical systems, 
there still remains rich phenomena that cannot be described 
with such typical oscillators.

Now, 
we present a new class of coupled oscillators 
in which  each oscillator has a heteroclinic cycle attractor\cite{rf:GH83}
instead of a limit cycle or a phase oscillator. 
Being different from those typical oscillators, 
an oscillator with a heteroclinic cycle attractor 
has no characteristic time scale by the following reason. 
A heteroclinic cycle is constructed with some saddle fixed points and 
heteroclinic orbits that connect the fixed points cyclically.
When an orbit approaches to a heteroclinic cycle attractor, 
it stays long in the neighborhood of fixed points and moves along with 
heteroclinic orbits quickly. 
The length of the staying time grows exponentially for each oscillation
while the moving speed between fixed points has little change. 
Then the period of oscillation gets exponentially longer 
and the system has no characteristic time scale. 
This property of heteroclinic cycles can bring quite complex 
structures such as the coexistence of infinitely many attractors\cite{rf:C97}.
%
Comparing with other oscillators with natural frequencies, 
this class of systems will give another plentiful phenomena 
and make our understandings of high dimensional dynamical systems richer.

\begin{figure}
\begin{center}
\includegraphics[width=10cm]{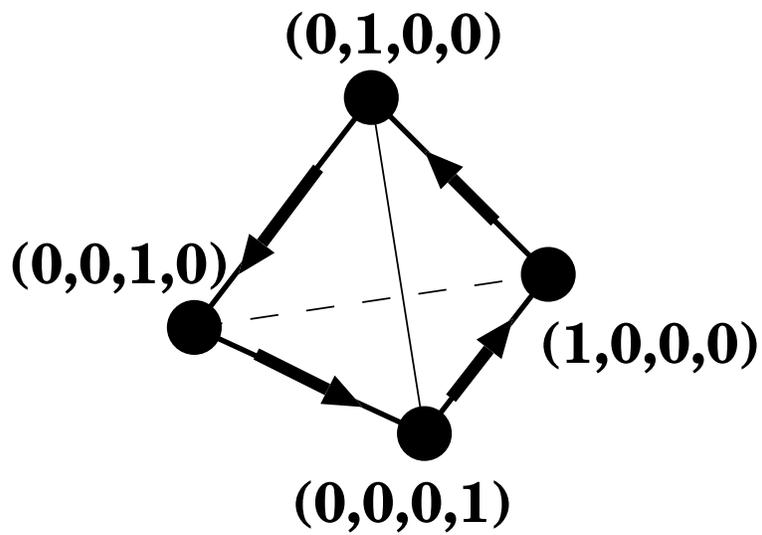}
\end{center}

     \caption{
A phase space of replicator system with 4 components and 
heteroclinic cycle. Filled circles are saddle fixed points and 
Thick solid lines with arrows are heteroclinic orbits.
}
\label{fig:1}
        \end{figure}

In a general dynamical systems, heteroclinic cycles are always 
structurally unstable\cite{rf:GH83}. 
However, if a system has certain constraints or symmetries, 
they can make invariant sets in the phase space and  
a robust heteroclinic cycle can exist in the invariant sets. 
As a system with such constraints, 
we adopt a replicator system\cite{rf:C97,rf:ML75,rf:B94,rf:C95,rf:HS98}
 for each oscillator, given by 
{\small
\begin{eqnarray}
\frac{dx_i}{dt} = 
x_i \bigl(\sum_{j=1}^4 a_{ij}x_j  -
\sum_{j,k=1}^4 a_{jk} x_j x_k\bigl), 
~~~~~~~~i=1,\ldots ,4 \label{eq:1}\\
\sum_{i=1}^4x_i=1,~~~~~0 \le x_i \le 1
~~~~~~~~~~~~~~~~~~\label{eq:2}
\end{eqnarray}
}
with the parameter matrix 
{\small
\begin{eqnarray}
(a_{ij})=\left(
\begin{array}{cccc}
 0 & -2 & -1 &  1 \\
 1 &  0 & -2 & -1 \\
-1 &  1 &  0 & -2 \\
-2 & -1 &  1 &  0 \\
\end{array}
\right).\label{eq:3}
\end{eqnarray}
}
With the constraints (\ref{eq:2}),
the phase space of the replicator system with 4 components 
is restricted in a tetrahedron 
with $\mbox{\boldmath $x$}=(1,0,0,0),(0,1,0,0), 
(0,0,1,0),(0,0,0,1)$ for the vertices(Fig.\ref{fig:1}).
Vertices, edges ({\small $\{\mbox{\boldmath $x$}|x_i+x_j=1,~ x_k=x_l=0 \}$}) 
and surfaces ({\small$\{\mbox{\boldmath $x$}|x_i+x_j+x_k=1,~x_l=0 \}$})
of the tetrahedron become invariant sets. 
In particular, vertices are always fixed points with any parameter matrix. 
Using the parameter matrix (\ref{eq:3}), 
an attracting heteroclinic cycle 
is constructed with the 4 vertices as saddle fixed points
and the cyclically connecting edges as heteroclinic orbits(Fig.\ref{fig:1}).

In our model, 
the replicator systems are distributed on a square lattice and 
coupled between nearest neighbors diffusively. 
The equations are given as 
{\small
\begin{eqnarray}
\frac{dx_i}{dt}^{(u,v)} &=& 
x^{(u,v)}_i \bigl(\sum_{j=1}^4 a_{ij}x^{(u,v)}_j  -
\sum_{j,k=1}^4 a_{jk} x^{(u,v)}_j x^{(u,v)}_k\bigl) \nonumber \\
&+& 
D(x^{(u-1,v)}_i+x^{(u,v-1)}_i+x^{(u+1,v)}_i+x^{(u,v+1)}_i- 4 x^{(u,v)}_i) ,
\label{eq:4}  \\
& &~~~~~~~~~~~~~~~~~~~~~~~~~~~~~
~~~~~~~~~~~~~~~~~~~~~~~~~~~~~~~~~~~i=1,\ldots ,4. \nonumber
\end{eqnarray}
}
where $(u,v)$ is a site index that stands for a location of an oscillator and 
$D$ is a diffusion constant between adjoining sites.
Free boundary condition is employed.
Since there is no non-diagonal diffusion term in the coupling method and 
the diffusion constant of each component has the same positive value, 
the spatially uniform oscillation never become unstable.

A replicator system is a model for a ecological system or a 
chemical reaction network of self-catalyzing molecules. 
Therefore the situation of 
our system can occur in a spotted ecological system with diffusion
(ex. a system on trees of an orchard) or a 
population dynamics of self-catalyzing proteins in cells.


Before we state our main results for a system on 2-dimensional arrays
we mention briefly about our preliminary results on 
systems with 1-dimensional lattices for comparison. 
If we choose free boundary condition for 1-dimensional systems
(number of oscillators $\ge 2$)
all replicators are synchronized 
and only the spatially uniform oscillation become the attractor.
Since all replicators are not chaotic and we choose 
the simple diffusive coupling as the interactions between oscillators
mentioned above, this result is easily acceptable.
If a system has periodic boundary (number of oscillators $\ge 3$)
 with a certain small diffusion constant, 
traveling waves as well as the spatially 
uniform oscillation become attractors. 

\begin{figure}
\begin{center}
\includegraphics[width=10cm]{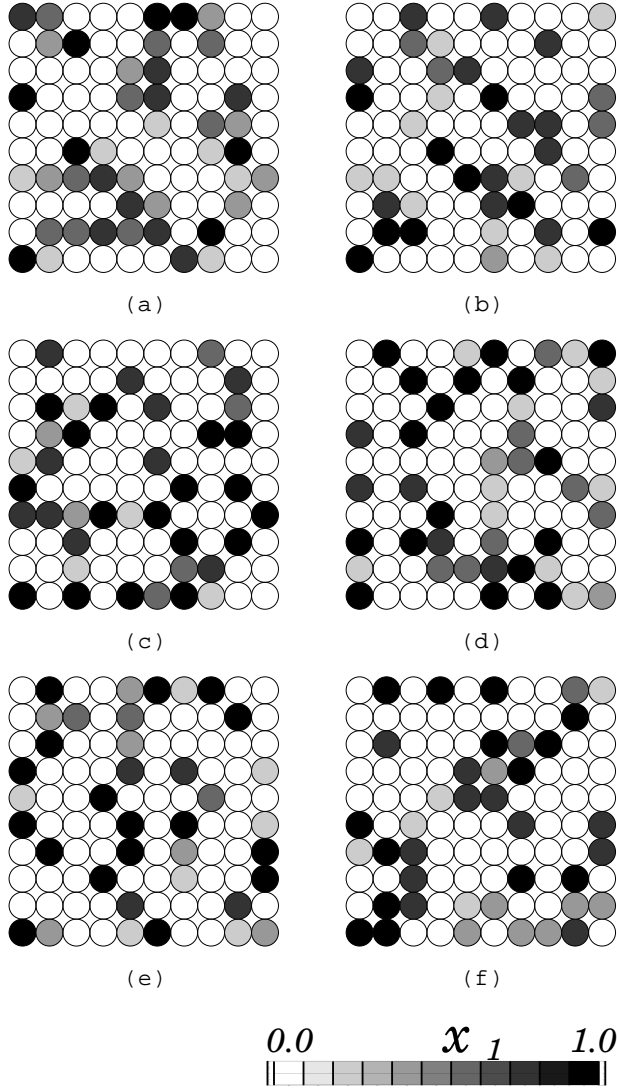}
\end{center}
           \caption{
6 snapshots of distributions of $x_1^{(u,v)}$ with a gray scale, 
$10 \times 10$ sites, $D = 10^{-4}$. 
(a)$\sim$(e) are generated from different initial conditions 
and taken the shots when the $(1,1)$-oscillators (the bottom left site) 
come to a same phase.
}\label{fig:2}
\end{figure}
\begin{figure}
\begin{center}
\includegraphics[width=10cm]{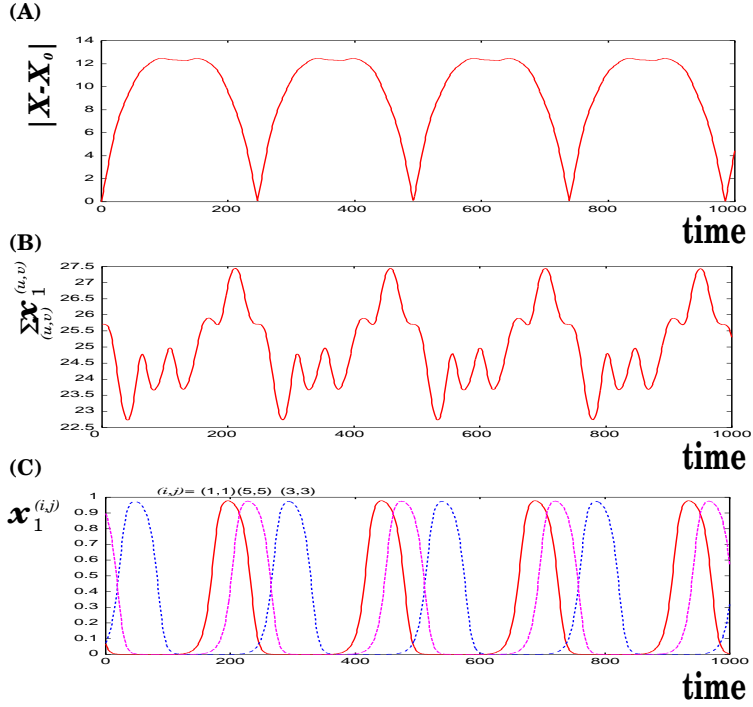}
\end{center}

     \caption{
Time series data of Fig.\ref{fig:1}-(e). 
(A): Distance between  $\mbox{\boldmath $X$}_0$
 and  $\mbox{\boldmath $X$}$ in the ambient phase space.
$\mbox{\boldmath $X$}_0=(x_1^{(1,1)}(t_0),x_2^{(1,1)}(t_0),$
$\ldots,x_3^{(10,10)}(t_0),x_4^{(10,10)}(t_0)) $  
is a point in a trajectory at a time $t_0$ (after approaching 
sufficiently close to an attractor) 
and $\mbox{\boldmath $X$}$ is the trajectory after the time.
(B): Mean field oscillation of $x_1$. 
(C): Oscillation of $x_1$ at (1,1),(3,3) and (5,5) sites. 
The exact recurrence in (A) indicate that the disordered pattern 
forms a limit cycle in the ambient phase space. 
Comparing (A) and (C), 
the oscillators synchronize with the limit cycle. 
}\label{fig:3}
     \end{figure}

In this letter, we report the discovery of a novel class of patterns 
that are spatially disordered but periodic in time 
(Fig.\ref{fig:2},\ref{fig:3}).  
From different initial conditions, 
a large variety of patterns are observed (Fig.\ref{fig:2}).  
Fig.\ref{fig:3} 
shows that these disordered patterns are stable limit cycle attractors 
in the ambient phase space. 
Hence, these patterns are exactly recurrent and not chaotic. 
The different patterns correspond to the different limit cycles 
and they divide the phase space as their basins.
After a system approaches sufficiently close to an attractor, 
frequency of each oscillator 
synchronizes with the limit cycle(Fig.\ref{fig:3}). 
This means that the frequencies of all oscillators are entrained 
while the phases of oscillations keep difference.
The disordered patterns are observed in the weak coupling range
($D \le 0.01$), 
while rotating spiral patterns of a well known type 
are seen with a certain large diffusion  constant ($D\simeq 0.1$).

To Understand the structures of the disordered  patterns, 
we note the local 
phase differences among oscillators. 
One can project the phase space of each replicator system
on a 2-dimensional plane 
($\alpha$-$\beta$ plane), 
\begin{eqnarray}
\begin{array}{l}
\alpha^{(u,v)} = x^{(u,v)}_1 - x^{(u,v)}_3\\
\beta^{(u,v)}  = x^{(u,v)}_2 - x^{(u,v)}_4
\end{array}.\label{eq:5}
\end{eqnarray}
The vertices are projected on $(\alpha,\beta)=(1,0),(0,1),(-1,0),(0,-1)$  
respectively.
We plot orbit-points ($\mbox{\boldmath $x$}^{(u,v)}(t)$) 
of all oscillators on 
the $\alpha$-$\beta$ plane 
and connect between coupled oscillators with lines.

\begin{figure}
\begin{center}
\includegraphics[width=10cm]{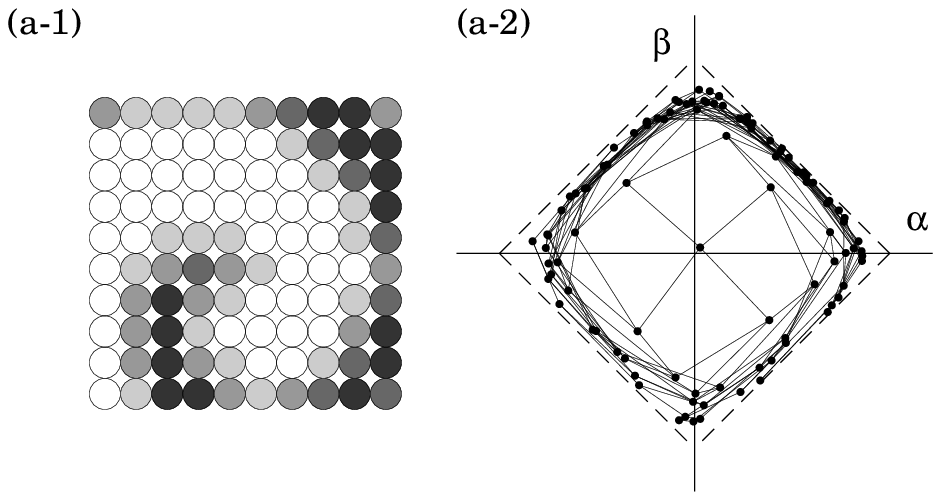}\\
\includegraphics[width=10cm]{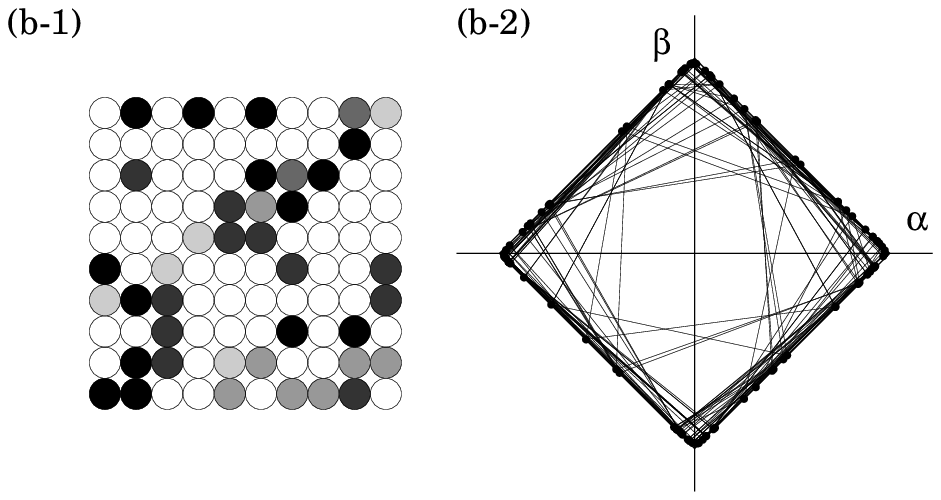}
\end{center}
     \caption{
The snapshots of $x_1^{(u,v)}$ distributions (a-1,b-1)and 
corresponding phase differences in $\alpha$-$\beta$ plane (a-2,b-2). 
In (a-2,b-2), dots indicate projected orbit-points and 
the couplings between them are presented by lines. 
The broken line in (a-2) means projected heteroclinic cycle.
$D=10^{-2}$ in (a)-system  and rotating spiral pattern 
is seen in (a-1).
$D=10^{-4}$ in (b)-system; our target pattern.  
}
\label{fig:4}
     \end{figure}

\begin{figure}
\begin{center}
\includegraphics[width=10cm]{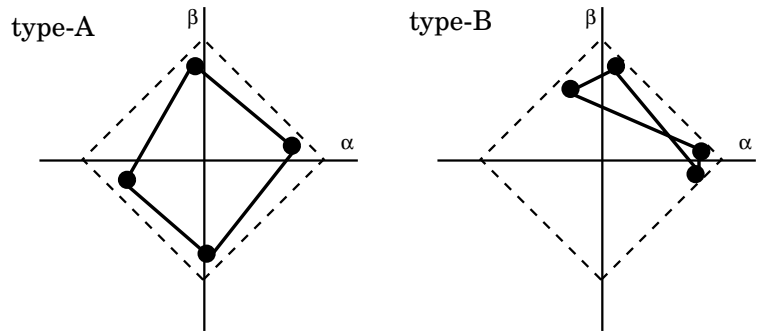}
\end{center}
     \caption{
Typical configurations of a set of 4 oscillators placed in a unit square.
}
\label{fig:5}
     \end{figure}

Typical snapshots after systems approach sufficiently close to attractors 
are shown in Fig.\ref{fig:4}. 
Comparing with Fig.\ref{fig:4}-a with a large $D$ , 
Fig.\ref{fig:4}-b displays 2 features as local laws among oscillators. 
First, 
all orbit-points stay close to the heteroclinic cycle. 
Second, distances between coupling lines and the origin of 
the $\alpha$-$\beta$ plane are not smaller than a certain positive value.
They are observed in every single step of numerical integrations. 
Therefore, the relation on locations of orbit-points 
for every two neighboring oscillators 
never become anti-phase.  
Considering these features, 
we conclude that 
the orbit-point arrangement of every set of the 4 oscillators 
that makes a unit square of the lattice (represented by dual-lattice point)
is classified into 2 types; orbit-points of 4 oscillators 
surround the center (type-A) or not (type-B).
Fig.\ref{fig:5} shows configurations of the types.

To investigate the difference of the types in detail, 
let us consider a system within $2 \times 2$ oscillators as the simplest case.
This system is thought as a system 
with 4 oscillators on 1-dimensional periodic array.
Therefore,  the situation of type-A corresponds to 
the traveling wave solution. It is an attractor when the diffusion constant
is smaller than about 0.01.
Because phases of oscillators are different from each other,
the diffusion effect prevents the orbit-points approaching to 
the heteroclinic cycle and the system keeps an uniform frequency.
In contrast, if the system is arranged in type-B as the initial condition,
oscillators are synchronized each other and 
the uniform oscillation approaches to the heteroclinic cycle.  
Therefore this system need to be arranged as type-A 
for the oscillation with an uniform frequency.

Returning to consider a large system, 
we note only on the type-As which generate oscillation
with an uniform frequency.
In such system, the type-As distributed on dual-lattice points
can be regarded as another type of vortices whose properties are 
different from those of the spiral pattern in Fig.\ref{fig:4}-a.
Therefore the disordered patterns are understood
as coexistences of many vortices and 
the distributions of vortices on dual-lattice points 
would be important on characterization of the patterns.

To make our characterization of the patterns clearer and more systematic, 
we note another property of this system.
As mentioned above, the diffusion effect keep orbit-points
from approaching to heteroclinic cycle.
Thus smaller diffusion constant makes 
orbit-points stay closer to the heteroclinic cycle and 
longer in neighborhoods of fixed points. 
Therefore, 
the time in a orbit-point passing along with a heteroclinic orbit 
become negligible 
and we see all orbit-points stay neighborhoods of the fixed points
in typical snap-shots.
With such small diffusion constant, 
we can systematically generate all possible snap-shots
as initial conditions
by choosing  all orbit-points of replicators 
in neighborhoods of any fixed points.

\begin{figure}
\begin{center}
\includegraphics[width=10cm]{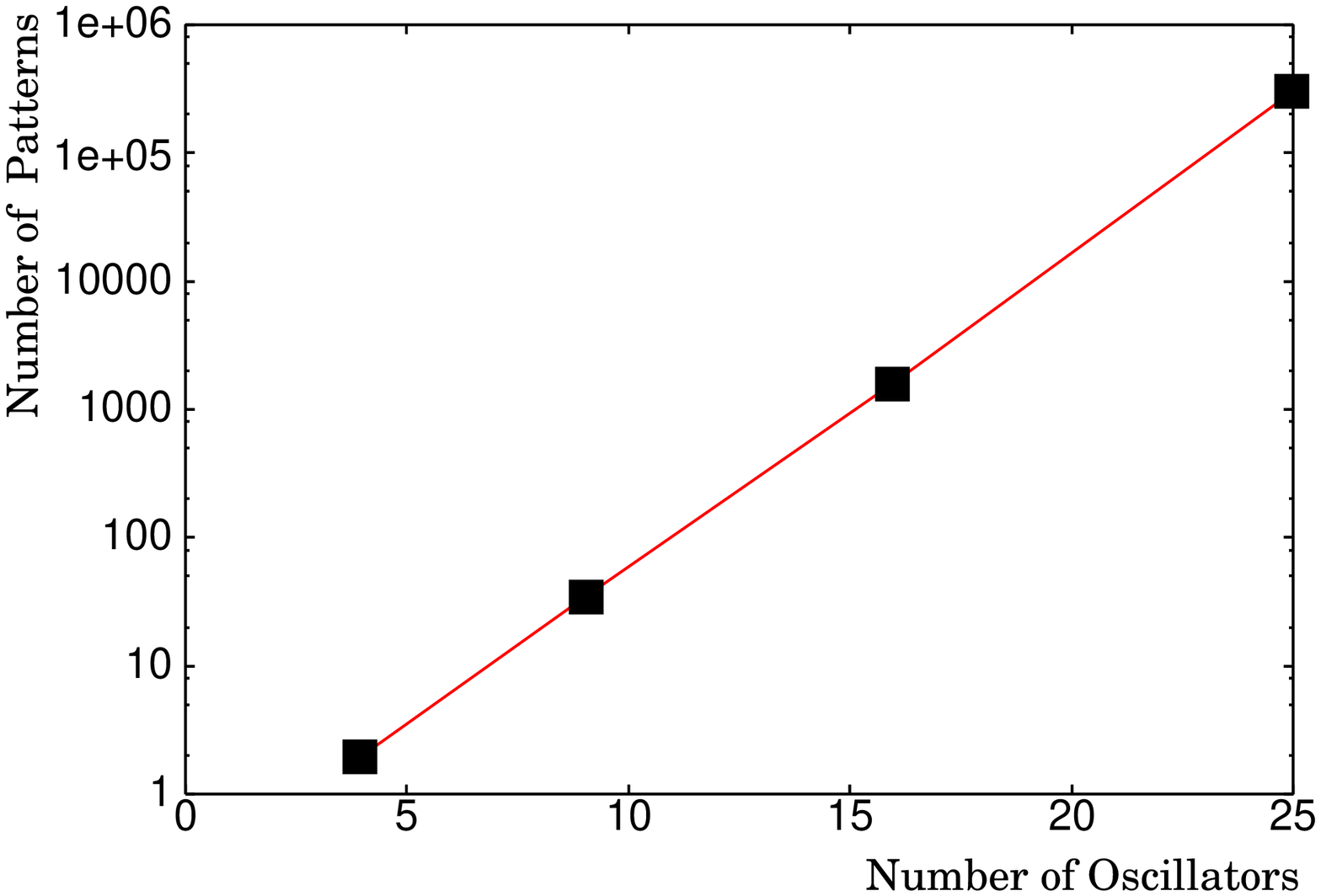}
\end{center}

     \caption{
The estimates of  the number of disordered patterns  
algorithmically calculated with condition (i).
Regular square case (2$\times$2, 3$\times$3, 4$\times$4 and 5$\times$5 
sites) are plotted.
}

\label{fig:6}

\end{figure}

Now, we choose the procedure for algorithmic generation 
of possible patterns as follows 
(i) distribute vortices (type-As) on the dual-lattice points without conflicts,
(ii) select a initial condition orbit-points of oscillators from
neighborhoods of 4 fixed points to satisfy  condition (i). 
We check all possible patterns for $3 \times 3$ and $4 \times 4$ oscillators 
systems numerically and find that 
every candidate which satisfies condition (i) 
have at least one corresponding attractor 
and most of them have only one attractor.
It means that almost all patterns are uniquely
characterized by the distributions of 
vortices.
Therefore we can estimate 
the number of attracting patterns with condition (i).
Fig.\ref{fig:6} shows the estimate 
that is algorithmically counted by computer. 
Since the condition (i) is a  combination problem, 
the number of patterns  grows exponentially with the increase of the  
number of oscillators. 

In this letter, 
we have reported spatially disordered oscillating patterns 
in the coupled heteroclinic cycles.
They have been shown periodic in time and not chaotic. 
The local law of differences of oscillators in the patterns
and the estimation of the number of stable patterns
have been also reported.
In our forthcoming paper\cite{rf:T02}, we will analyze periods and stability 
of the patterns in detail.

Remarkable point of this system is that there is no obvious effect 
for violations from spatially 
uniform oscillation such as chaotic nature of each 
oscillator, variety of frequencies or anomalous coupling. 
Therefore the uniform oscillation is always an attractor. 
Nevertheless, disordered patterns can exist 
as coexistence of vortices which structure is different from 
the well-known spiral pattern. 
This type of disordered oscillating patterns 
has not been found in
any other coupled oscillators.
However, 
other types of  spatially disordered stable patterns have been reported
in some class of dynamical systems\cite{rf:D97,rf:FEY93,rf:FESY95,rf:CMV96}. 
The relations among these patterns and ours will be discussed 
in the paper\cite{rf:T02}.

I would like to thank T. Konishi, H. Yamada, and S. Sasa for 
discussions and useful advises.

\end{document}